# Optical Pre-Emphasis by Cascaded Graphene Electro Absorption Modulators

V. Sorianello, G. Contestabile, M. Midrio, M. Pantouvaki, I. Asselbergs, J. Van Campenhout, C. Huyghebaerts, and M. Romagnoli

*Abstract*—A simple optical circuit made by a cascade of two graphene-on-silicon electro absorption modulators (EAMs) of different length is used for the optical pre-emphasis of 10 Gb/s non-return-to-zero (NRZ) signals by delay-inverse-weight compensation. Transmission up to 100 km on single mode fiber (SMF) without dispersion compensation is reported, showing also the large performance advantage (6 dB in back-to back and around 5 dB in transmission) in respect of the conventional single EAM transmitter configuration.

## I. Introduction

GRAPHENE is nowadays extensively studied for a large range of applications including its use in photonic integrated circuits (PICs) [1]. In PICs, graphene is expected to improve device performance simplifying, at the same time, the fabrication technology [2]. In particular, graphene-based photodetectors do not need Ge epitaxy (currently used for Si photonics photodetectors), and also work on much larger spectral regions. Moreover, they can be extremely fast, and work at zero dark current when operated in voltage-detection mode [3], [4]. Similarly, graphene-based electro-optic modulators have a number of advantages over Si-based modulators. They are capable of efficient and broadband electro-absorption or phase modulation and are compatible with complementary metal oxide semiconductor (CMOS) processing with simplified post-processing fabrication on various substrates. Indeed, graphene modulators do not strictly require Si or Ge doping, hence, Si, SiN, $SiO_2$ or other materials can be used as waveguides [2] and short devices (of the order of 50-150 microns length), driven in lumped mode, can be realized.

Recently, broadband electro-absorption [5] and pure phase modulation [6] have been reported on a Si platform using short single layer graphene-based modulators. Graphene electro absorption modulators (EAMs), in particular, have been reported working at 10 Gb/s [7], and very recently, at 20 Gb/s [8], using a single layer configuration, being mainly bandwidth limited by the graphene-metal electrical contact resistance value. Additionally, graphene EAMs show significant positive linear chirp associated to the Fermi level electrical gating responsible for the absorption changes [9]. The presence of chirp with the right sign on the modulated signals makes possible chromatic dispersion compensation while transmitting on standard single mode fibers (SMFs). Indeed, the combination of linearly chirped signals and anomalous dispersion in SMFs counteracts the signal spreading due to the fiber chromatic dispersion. This pulse temporal lensing effect make possible the transmission of 10 Gb/s signals on, at least, 100 km SMF, as reported in [9].

In this letter, we show that signals generated by graphene EAMs can be significantly improved by optical pre-emphasis using a simple compact circuit made cascading two short graphene EAMs driven by two proper complementary electrical signals. This paper is an extended version of a recently presented conference paper [10], and reports 10 Gb/s back-to-back and transmission results obtained with the double EAM modulator, also performing a comparison with the single modulator case.

## II. Device Description

The fabricated device is made by cascading two single layer Graphene (SLG) on Si waveguide EAMs similar to the one recently reported in [7], [9]. As shown in Fig. 1, it is made by input/output vertical coupling gratings and the cascade of a 100 and a 50 $\mu$m-long EAM.

The modulators are made by Si photonic waveguides with SLG transferred on top of the Si waveguide core. The Si photonic part was realized within the IMEC iSiPP25G silicon on insulator (SOI) platform [11]. The Si ridge waveguide is designed to support a single transverse electric (TE) in-plane polarized optical mode with a core cross-section of 480 nm x 220 nm and 60 nm slab. The waveguide is boron doped to reduce the Si electrode resistance and allow high-speed operation. The waveguide $SiO_2$ top cladding is thinned down to 10 nm on the top of the waveguide core. SLG is grown

Manuscript received February 15, 2019; revised April 17, 2019; accepted April 24, 2019. Date of publication May 2, 2019; date of current version May 23, 2019. This work was supported in part by the Project GrapheneCore1 through the European Commission under Grant 696656. *(Corresponding author: G. Contestabile.)*

V. Sorianello and M. Romagnoli are with the Photonic Networks and Technologies Laboratory, Consorzio Nazionale per le Telecomunicazioni (CNIT), 56124 Pisa, Italy.

G. Contestabile is with the Photonic Networks and Technologies Laboratory, Consorzio Nazionale per le Telecomunicazioni (CNIT), 56124 Pisa, Italy, and also with Scuola Superiore Sant'Anna, TeCIP Institute, 56124 Pisa, Italy (e-mail: contesta@sssup.it).

M. Midrio is with the Photonic Networks and Technologies Laboratory, Consorzio Nazionale per le Telecomunicazioni (CNIT), 56124 Pisa, Italy, and also with the Engineering Department, University of Udine, 33100 Udine, Italy.

M. Pantouvaki, I. Asselbergs, J. Van Campenhout, and C. Huyghebaerts are with the IMEC, 3001 Leuven, Belgium.

Color versions of one or more of the figures in this letter are available online at http://ieeexplore.ieee.org.

Digital Object Identifier 10.1109/LPT.2019.2914366

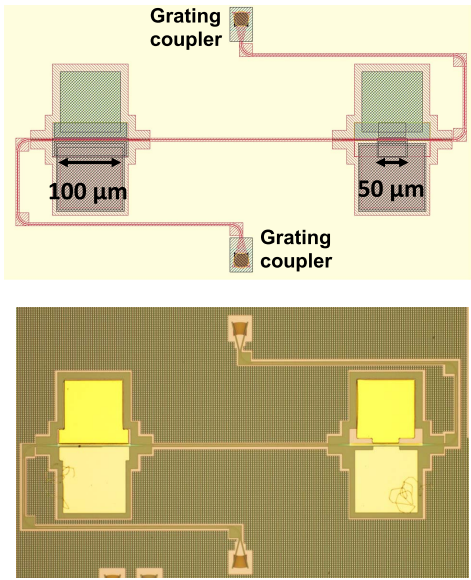

Fig. 1. Top part, mask design of the cascaded graphene EAMs. Lower part, microscope picture of a fabricated sample.

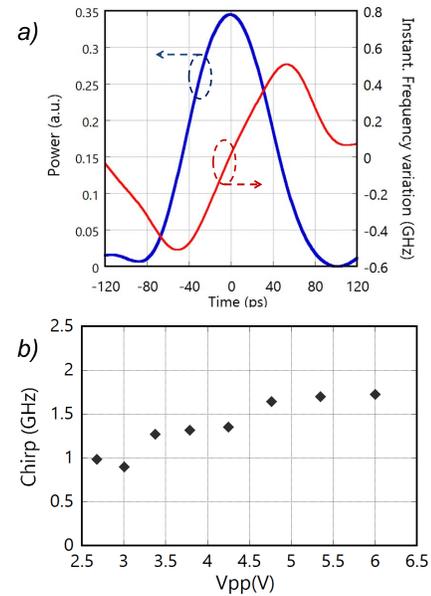

Fig. 2. a) Pulse amplitude profile (blue line) and measured instantaneous frequency variation (red line) versus time for signals generated by the graphene EAM. b) Pulse chirp dependence from the driving voltage.

by chemical vapor deposition (CVD) on copper foil [12], and then transferred on the waveguide by standard wet transfer [13]. Graphene was shaped through optical lithography and oxygen ashing. Metal contacts were placed by metal lift-off on the SLG and Si with separate processes: Palladium (Pd) is used for the SLG, while Si is contacted with a stack of Titanium (Ti)/Platinum (Pt)/Gold (Au). The 10 nm-thick layer of $SiO_2$ insulates Graphene from the Si waveguide core and forms a Silicon-Insulator-Graphene (SIG) capacitor [14]. Fermi Level on Graphene may be changed by applying a voltage across the SIG capacitor. Standard single polarization grating couplers having 5 dB insertion loss at 1550 nm are also included in the device design for input and output vertical fiber coupling.

As recently reported in [9], the 100 $\mu$m graphene EAM has a 5 GHz 3-dB electrooptic bandwidth, a modulation efficiency of 1dB/V, a maximum attainable extinction ratio of 4.5 dB. In addition, itshows an opposite optical transmissivity and refractive index change dependence with respect to the applied voltage. This gives, in a certain range of Fermi levels, an instantaneous and linear frequency chirp on the carved optical signal as shown in Fig. 2a. The figure reports the signal chirp in GHz measured by a complex spectrum analyzer in correspondence of an isolated signal pulse. Moreover, it results that the amount of linear chirp depends on the applied driving voltage as reported in Fig. 2b.

This modulator pre-chirping can be used for fiber chromatic dispersion compensation through the time lens effect and, thanks to this effect, 100 km transmission on SMF at 10 Gb/s has been reported [9] demonstrating a self-focusing distance (transmission distance for which there is the same signal sensitivity as in back-to-back) of 60 km (at a driving voltage of 2.5 Vpp).

This chromatic dispersion resilience can be combined with the optical pre-emphasis obtained by cascading two EAMs: indeed, when the cascaded circuit is driven with data, and properly attenuated and delayed inverted-data signals, the optical circuit emulates a 1-tap filter frequency high-pass circuit similar to the one reported in Fig. 3. This circuit performs delay inverse weight compensation: the effect of the delay, inversion, and weight, due to the second modulator, is to compensate for the slower part of the response of the first modulator (so enhancing the slow signal transitions due to the bandwidth limitation).

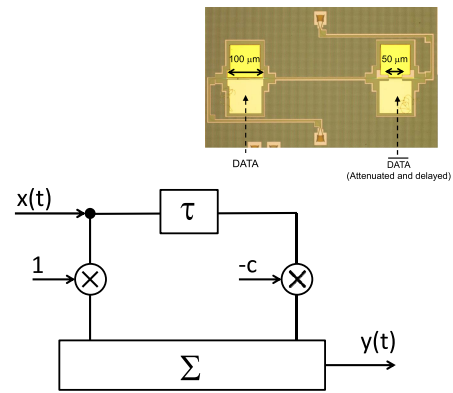

Fig. 3. Scheme of the electrical equivalent 1-tap pre-emphasis circuit which approximates the cascaded modulators when electrically driven like in the picture in the inset.

III. EXPERIMENTAL RESULTS

The setup for the transmission experiment is reported in Fig. 4. An optical continuous wave (CW) signal generated by an external cavity laser (ECL) at 1550 nm was coupled into the chip through a TE vertical grating coupler. The two EAMs were driven by 10 Gb/s data and inverted data through two Bias Tees. A $2^{31}-1$ long non-return-to-zero (NRZ) pseudo random bit sequence (PRBS) with a peak-to-peak voltage of 2.5Vpp was sent to the first EAM and the same delayed and inverted sequence with a peak-to-peak voltage of 0.8 V to the second one. Signal time delay between the two electrical signals was 32 ps. The EAMs were biased at −1.3 V. Chip insertion loss was about 19 dB so distributed: 10 dB due to the grating couplers, 4 dB (estimated by simulations) due to the two modulators, 5 dB due to scattering loss of the waveguide. Those 5 dB extra losses are due to the actual fabrication process [6], [7] which leaves the Si waveguides with a suboptimal cladding layer.

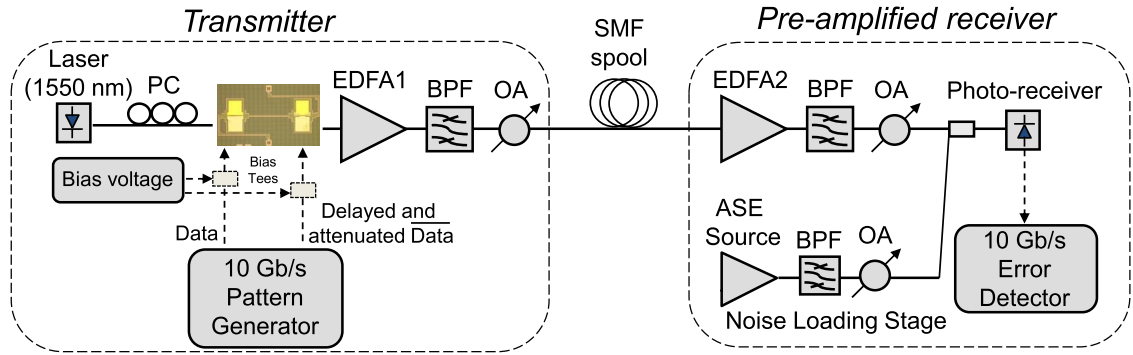

Fig. 4. Experimental set-up. PC= Polarization Controller, BPF= Optical Band-Pass Filter, OA= Optical Attenuator; EDFA= Erbium Doped Fiber Amplifier.

An erbium doped fiber amplifier (EDFA) was used to compensate for the chip losses and an optical band pass filter (BPF) with a bandwidth of 1 nm to remove the out-band optical noise introduced by the amplifier. A variable optical attenuator (OA) was also used to control the optical power at the input of the transmission fiber. SMF spools were of different length, from 10 to 100 km, all having a dispersion parameter D=17 ps/nm/km ($\beta_2 = -21.6$ ps$^2$/km). At the output of the SMF, another set of EDFA, BPF and OA was used to compensate the fiber propagation losses. A noise loading stage based on an amplified spontaneous emission (ASE) source, a BPF and OA, was used to properly perform bit error rate (BER) measurement versus signal ASE. The noise loaded optical signal was collected with a photo-receiver and analyzed with an error detector.

The effect of the optical pre-emphasis by the cascaded EAMs can be clearly seen starting from the back-to-back eye diagrams reported in Fig. 5. On the right top part, the best attainable eye diagram with the single EAM, on the lower part, the best one by the cascaded EAMs.

The comparison is made using a 100 $\mu$m long single EAM, like the one reported in [9], and the 100 $\mu$m plus the 50 $\mu$m long modulator driven like reported in Fig. 3, i.e., with the same sequence with opposite sign, after attenuation and proper delay. The main modulating device is still the 100 $\mu$m long EAM with the same absorption length, series resistance, capacitance, and insertion loss, the second device contributes with the slight over-modulation needed to compensate for the slower part of the modulator response, at the expense of an extra loss.

In both cases the extinction ratio is 2.5 dB, but, the cascaded eye diagram is significantly more clear and open with a net reduction of the pattern sequence dependence. This is due to the enhanced digital transitions by high pass optical filtering of the modulator cascade.

Overall, this gives a Q-factor improvement, measured by the electrical sampling scope, of around 2, from 3.4 to 5.3.

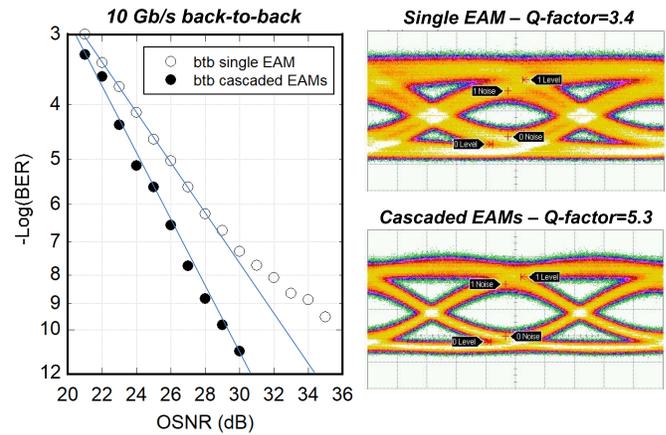

Fig. 5. Single and cascaded graphene EAM transmitter BER vs. OSNR in back-to-back with correspondent eye diagrams. Time reference = 20 ps/div. Blue linear trend curves are also added in the graph.

The oscilloscope measures the difference of the mean values of the two signal levels (level for a "1" bit and level for a "0" bit) divided by the sum of the noise standard deviations at the two signal levels.

Correspondent BER measurements, also reported in Fig. 5 (as a function of the optical signal to noise ratio measured on 0.1 nm), show an OSNR gain of 6 dB for the cascaded modulators and the cancellation of the slight floor tendency given by the modulator bandwidth limitations (EAM bandwidth ∼ 5 GHz [9]). The back-to-back sensitivity (at $10^{-9}$ BER) is at an OSNR of 28 dB. However, as reported in Fig. 6, and similarly to what happens to the single graphene EAM [9], the time lens effect, given by the combination of the signal chirp and the fiber dispersion, compresses the optical bits giving a net eye opening and a correspondent sensitivity improvement. The improvement has its maximum at 40 km, where the sensitivity is 27 dB (a slight eye opening can be also appreciated comparing the eye diagrams at 10 and 40 km reported in the right part of Fig.6). The self-focusing distance, the one for which there is a 0 dB OSNR penalty, is about 60 km. From there on, additional fiber propagation increases the transmission penalty because of the chromatic dispersion pulse broadening which increases the inter-symbol interference. In particular, at 100 km, the signal shows a floor tendency at $10^{-9}$. Nevertheless, longer transmission without chromatic dispersion compensation with a signal BER significantly lower than the conventional Reed-Solomon forward error correction (RS-FEC) threshold of $10^{-3}$ is clearly possible.

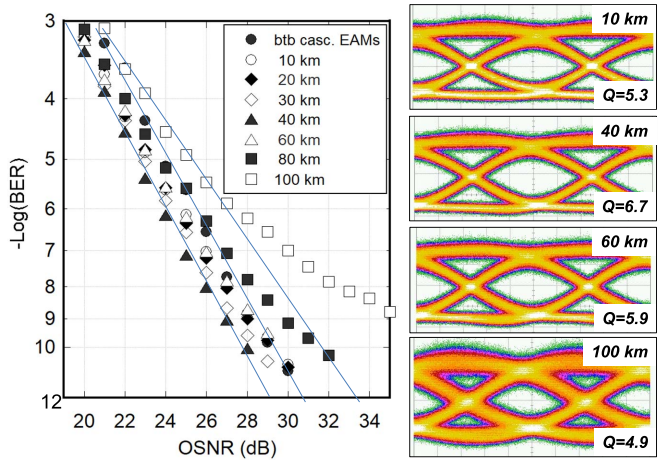

Fig. 6. Cascaded graphene EAM transmitter BER vs. OSNR for signals transmitted up to 100 km SMF. On the right, sample eye diagrams. Time reference = 20 ps/div. Blue linear trend curves are also added in the graph.

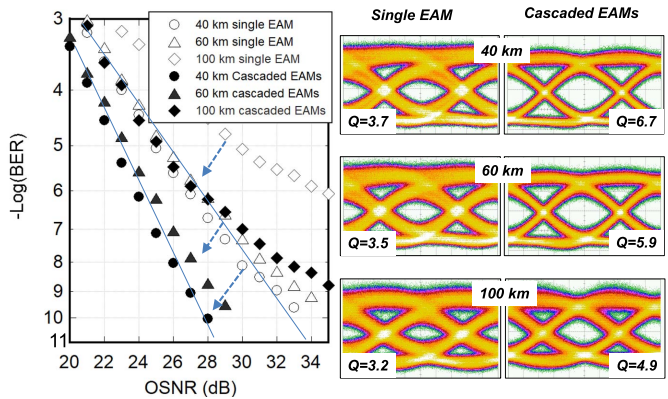

Fig. 7. Single and cascaded graphene EAM transmitter BER vs. OSNR at 40, 60 and 100 km with correspondent eye diagrams. Time reference = 20 ps/div. Blue linear trend curves are also added in the graph.

In order to compare these results with the single EAM, sample eye diagrams and BER curves of cascaded and single EAM (100 $\mu$m-long) are reported in Fig. 7, together with the corresponding BER curves for the cases 40, 60 and 100 km.

The eye opening gain at the transmitter, reported in Fig. 5, gives an improved transmitted signal at every distance so that the 6 dB OSNR advantage in back-to-back is still more than 5 dB at 40 and 60 km and also removes the BER floor at $10^{-6}$ at 100 km. The signal improvement for every of the transmission distances can be clearly envisioned by comparing the recorded eye diagrams for the two cases reported in the right part of Fig. 7.

## IV. Conclusion

A novel silicon-graphene circuit which implements the optical pre-emphasis for NRZ signals, made by the cascade of a 100 and a 50 $\mu$m long EAM, has been fabricated and reported, showing a significant improved sensitivity in back-to-back (6 dB) and in fiber transmission up to 100 km, in respect of the single graphene EAM. The delay inverse weight high pass filtering optical pre-processing is obtained by driving the second shorter modulator with a delayed inverted and attenuated copy of the modulating electrical signal. Despite 10 Gb/s operation has been reported in this letter with a 5 GHz bandwidth modulator, this is a very promising technique for obtaining much faster operation when larger bandwidth graphene EAM modulators are used.

## Acknowledgments

The authors would like to acknowledge Graphenea, Spain for the supply of the CVD-grown graphene samples.